\documentclass[a4paper,11pt]{article}
\pdfoutput=1 

\usepackage{jinstpub} 
\usepackage{graphicx}
\usepackage{subfigure}

\title{\boldmath An improved design of the readout base board of the photomultiplier tube for future PandaX dark matter experiments}

\author[a,c]{Qibin Zheng,}
\author[a]{Yanlin Huang,}
\author[b]{Di Huang,}
\author[b,d]{Jianglai Liu,}
\author[e]{Xiangxiang Ren,}
\author[e]{Anqing Wang,}
\author[e]{Meng Wang,}
\author[b]{Jijun Yang,}
\author[b]{Binbin Yan,}
\author[b]{Yong Yang} \note{Corresponding author.}

\affiliation[a]{Institute of Biomedical Engineering, University of Shanghai for Science and Technology, \\Shanghai 200093, China}
\affiliation[b]{INPAC, Department of Physics and Astronomy, Shanghai Jiao Tong University, \\Shanghai Laboratory for Particle Physics and Cosmology, Shanghai 200240, China}
\affiliation[c]{Terahertz Technology Innovation Research Institute, University of Shanghai for Science and Technology, \\Shanghai 200093, China}
\affiliation[d]{Tsung-Dao Lee Institute, Shanghai Jiaotong University, Shanghai, 200240, China}
\affiliation[e]{School of Physics and Key Laboratory of Particle Physics and ParticleIrradiation (MOE), Shandong University, Jinan 250100, China}

\emailAdd{Yong.Yang@sjtu.edu.cn}

\abstract{
  The PandaX project consists of a series of xenon-based experiments that are used to search for dark matter (DM) particles and to study the fundamental properties of neutrinos. The next DM experiment PandaX-4T will be using 4 ton liquid xenon in the sensitive volume, which is nearly a factor of seven larger than that of the previous experiment PandaX-II. Due to the increasing target mass, the sensitivity of searching for both DM and neutrinoless double-beta decay ($0\nu\beta\beta$) signals in the same detector will be significantly improved. However, the typical energy of interest for $0\nu\beta\beta$ signals is at the MeV scale, which is much higher than that of most popular DM signals. In the baseline readout scheme of the photomultiplier tubes (PMTs), the dynamic range is very limited. Signals from the majority of PMTs in the top array of the detector are heavily saturated at MeV energies. This deteriorates the $0\nu\beta\beta$ search sensitivity. In this paper we report a new design of the readout base board of the PMTs for future PandaX DM experiments and present its improved performance on the dynamic range. 
 }
\keywords{Photomultiplier Tube; Large Dynamic Range; PandaX.}

\begin{document}
\maketitle
\flushbottom

\section{Introduction}

In recent years, dark matter (DM) experiments with dual-phase xenon time project chamber (TPC) technique, such as PandaX~\cite{Cui:2017nnn}, LUX~\cite{Akerib:2016vxi} and XENON~\cite{Aprile:2018dbl}, have been leading the search for weakly interacting massive particles (WIMPs) with masses above $\sim$5 GeV/c$^2$. WIMPs are among the leading hypothetical particle physics candidates for DM. In these experiments, energy deposition in liquid xenon generates prompt scintillation photons (usually called S1 signal) and delayed electroluminescence photons (S2 signal) in gaseous xenon, which leads to excellent background suppression and signal-background event discrimination. The 3D position of the energy deposition can be reconstructed using S1 and S2 signals. This makes it possible to fiducialize the central region where most of the ambient radioactivity is absorbed by liquid xenon outside the fidiucial volume. In addition, nuclear recoil events from WIMP-xenon interaction can be discriminated from electron recoil events from $\gamma$ or $\beta$-xenon interaction using the ratio of S2 and S1 signals. These features together with the scalability of the detector keep this type of experiments in the forefront of searching for WIMPs. With increasing target mass of xenon from $\sim$10 kg to $\sim$1 ton, the WIMP search sensitivities of the above-mentioned three experiments have been improved by 2-3 orders of magnitude in the last decade. All these experiments are being upgraded to the next stage of the DM search with a multi-ton target. For example, PandaX-4T experiment~\cite{Zhang:pandax4T} will be using 4 ton liquid xenon as the detection medium.

With an increasing mass of liquid xenon in these experiments, the detector also becomes much more capable in studying the properties of neutrinos. For example, one can search for neutrinoless double-beta decay ($0\nu\beta\beta$) of $^{136}$Xe, given the fact that there is ~8.9\% of $^{136}$Xe in the natural xenon. Detection of $0\nu\beta\beta$ is a direct evidence of neutrinos being their own antiparticles and violation of lepton number conservation law, which are clear signs of new physics beyond the standard model of particle physics. Recently, PandaX-II collaboration published a first search result for $0\nu\beta\beta$ of $^{136}$Xe with 580 kg xenon~\cite{Ni:2019kms}. However, the obtained limit on $0\nu\beta\beta$ half-life (2.4$\times10^{23}$ yr) is several orders of magnitude weaker than the currently most stringent result ($1.1\times10^{26}$ yr) from KamLAND-Zen, a dedicated $0\nu\beta\beta$ experiment. This is mainly due to much higher background rate and limited performance of energy resolution at the MeV energy scale. PandaX-II was designed primarily for WIMP search. The typical deposited energy expected from WIMP-xenon interaction is less than 10 keV. However, the most region of interest for $0\nu\beta\beta$ signals is 2-3 MeV (the Q value of $\beta\beta$ decay of $^{136}$Xe is 2.48 MeV). In PandaX-II, the response of photomultiplier tubes (PMTs) in the top array for S2 signals start to become nonlinear at energies higher than $\sim$100 keV due to PMT saturation. This deteriorates both the energy scale and resolution significantly for energies above MeV and affects the $0\nu\beta\beta$ search sensitivity. From the study in~\cite{Ni:2019kms}, a typical S2 signal in the top PMT array from the energy deposition of a 2.6 MeV gamma from $^{208}$Tl corresponds to 1000k photoelectrons(PEs). The PMT with the maximum number of PEs contributes to about 30\%. In PandaX-4T, the width of the gaseous xenon region is expected to be reduced by half compared to PandaX-II since multiple scattering events can be better identified with narrower S2 signals. Therefore, for the purpose of searching for both DM and $0\nu\beta\beta$ signals in the same detector, the dynamic range of the PMT needs to be above 150k PE, which is much higher than that of the PMT in the baseline readout scheme in PandaX-4T.   

In this paper, we revise the baseline design of the readout base board used for PandaX-4T experiment to improve the PMT's dynamic range. The base board is a printed circuit board which not only hosts the PMT, but also provides high voltage (HV) for each dynode and the anode of the PMT. In the baseline design, only one signal can be read out from the anode. In the new design, one additional signal can be read out from the eighth dynode (DY8) with smaller gain but larger dynamic range. The rest of the paper is organized as follows. In section~\ref{design} we describe the  design of the new base board. In section~\ref{measurement} we describe the system of measuring the dynamic range of the PMT and present the results from both the baseline and the new base board. In section~\ref{discussion} the results are summarized.
  
\section{Design of the new base board}
\label{design}

\begin{figure*}[!htbp]
  \centering
  \subfigure[Baseline design of the base board]{
  \includegraphics[width=6in]{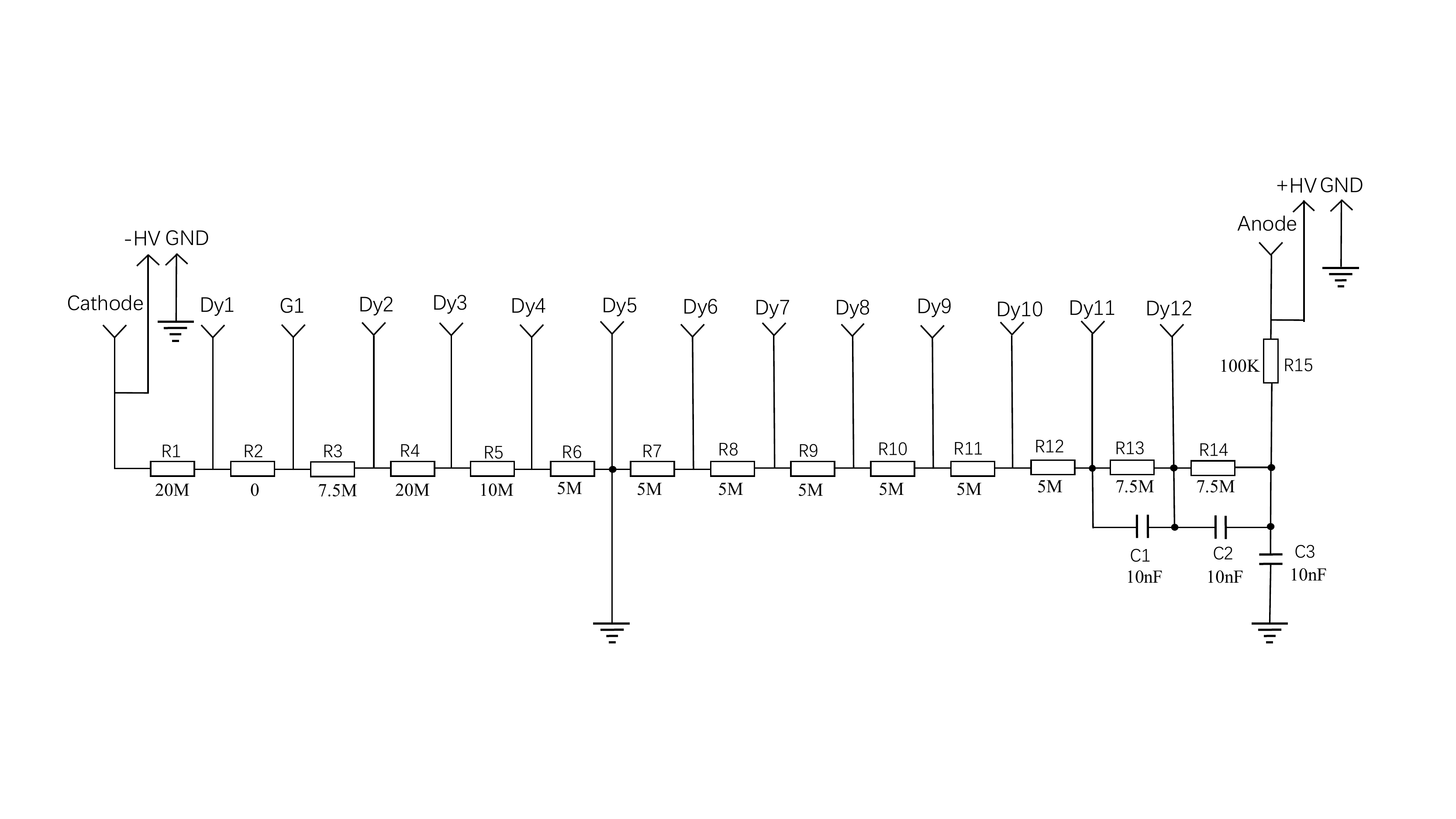}
  \label{two_design_a} }
  \hspace{1in}
  \subfigure[New design of the base board]{
  \includegraphics[width=6in]{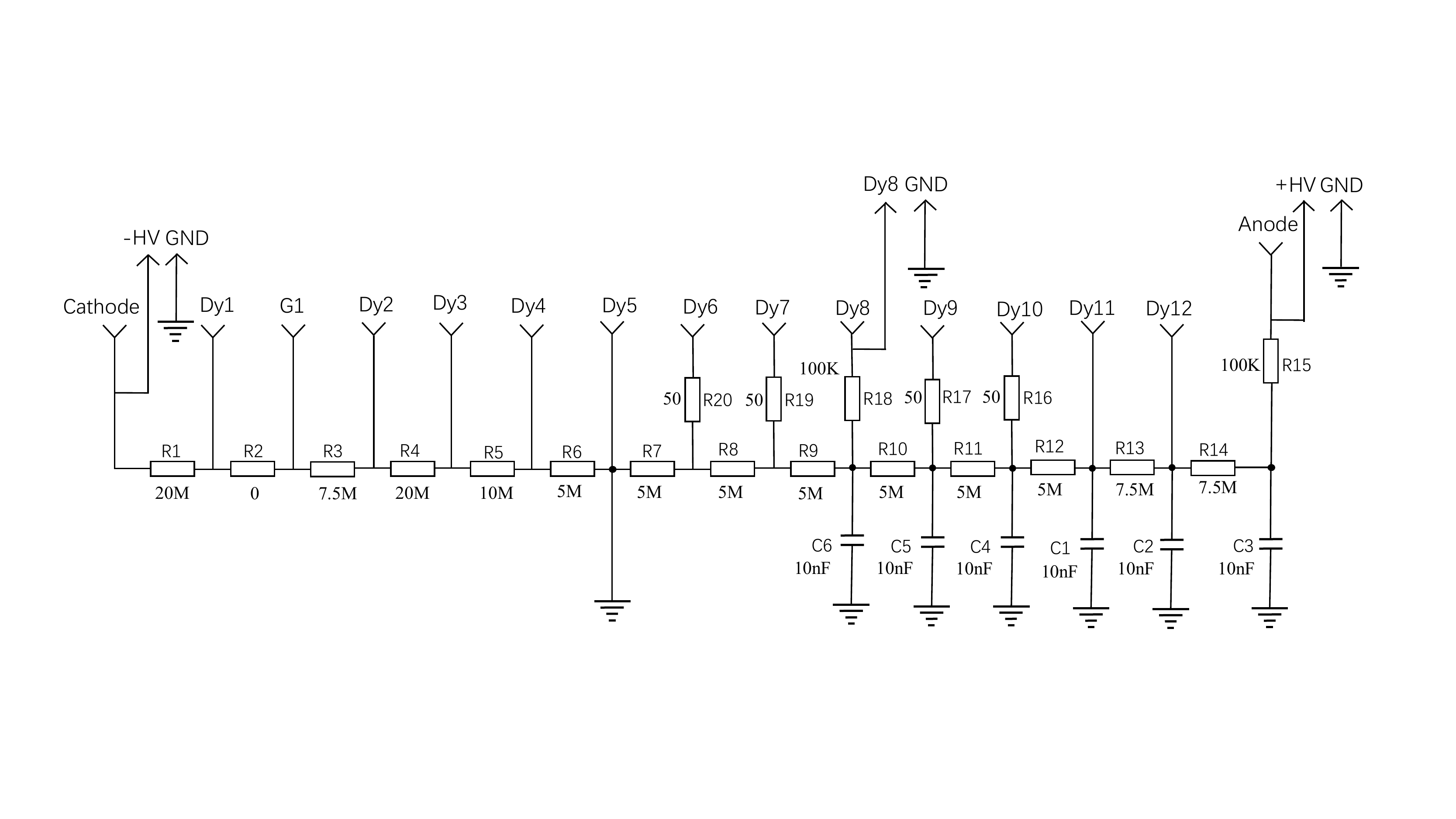}
  \label{two_design_b}
  }
  \caption{The schematics of baseline (a) and the new (b) base board. The main differences are the filtering capacitors configuration and the additional readout at DY8. For measuring the dynamic range, the cathode, dynode 5 and the anode were connected to -700 V, GND and +800 V, respectively.}
  \label{two_design}
\end{figure*}

Both experiments of PandaX-II and PandaX-4T choose Hamamatsu R11410-20 3-inch PMTs to detect scintillation photons. The typical quantum efficiency is approximately 30\% for the 175 nm xenon scintillation light. Photoelectrons at the cathode are amplified through 12 dynodes and the anode. The typical gain is 5$\times 10^{6}$ when the cathode and the anode is supplied with the HV of 1500 V. In PandaX-II and PandaX-4T, a split positive and negative HV scheme is adopted to reduce the relative potential to ground, which reduces the risk of discharge on the feedthrough pins. The voltage at each dynode is set through a resistor network shown in Figure~\ref{two_design_a}, which is the schematic of the baseline design of the base board for PandaX-4T. In the baseline design, PMT signal is read out from the anode. As suggested in~\cite{ref:hamamatsu}, usually a few decoupling capacitors are added at the last stages, which provides additional charge when the electrical pulse is being read out. This in turn improves the linearity of the PMT. Usually the decoupling capacitors are connected in series or parallel. Two capacitors (C$_{1}$,C$_{2}$) are connected in series at the last two dynodes. One additional capacitor (C$_{3}$) is required to be connected in series with the load resistor (R$_{15}$) of the anode since a positive HV is applied at the anode. 

To improve the dynamic range of the PMT, a few modifications were made based on the baseline design, as shown in Figure~\ref{two_design_b}. First is the way the decoupling capacitors are connected. In the new design, each capacitor at the dynode is connected to ground directly. As we will show later, adding three more capacitors from DY8 to DY10 connected in parallel to ground improves the dynamic range of the anode significantly. Secondly, to further improve the dynamic range, another electrical signal with smaller gain is read out from DY8 at the same time. And during the test, we observed some damped oscillation in the falling edge of the electrical pulse from DY8. As suggested in~\cite{ref:hamamatsu}, a few matching resistors are added near DY8 and the oscillation is found to be significantly reduced. 
 
\section{Dynamic range measurement}
\label{measurement} 

\begin{figure*}[!htbp]
\centering
\includegraphics[width=1.1\textwidth]{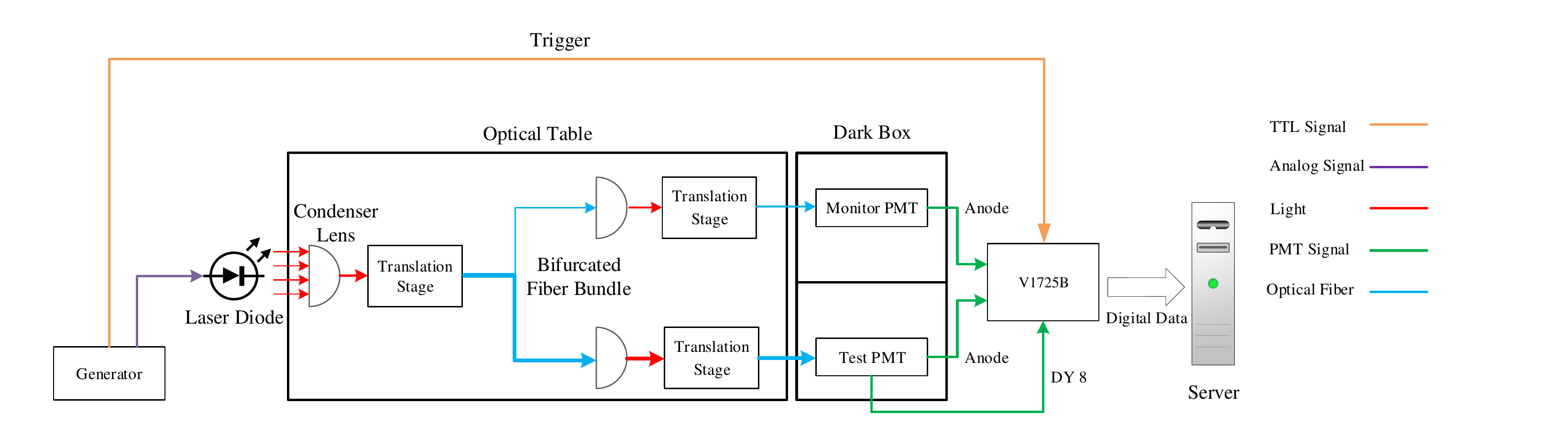}
\caption{The system for PMT dynamic range measurement. This consists mainly of a light source and splitting system, a dark box, a waveform digitizer and a DAQ server.}
\label{test_setup}
\end{figure*}

To measure the dynamic range of both the baseline and the new base boards, we set up a system shown in Figure~\ref{test_setup}. Driven by an arbitrary waveform generator~\cite{ref:generator}, the laser diode~\cite{ref:L450P1600MM} emits light with a fixed duration time and frequency. The intensity of the light can be adjusted such that the PMT under test can be saturated. For measuring the dynamic range, we set the duration time to be 10$\mu$s, which is the typical width of S2 signals at MeV energies expected in PandaX-4T. The frequency is set to be 50 Hz. To increase the light detection efficiency of the PMT, the light is focused by a focusing lens. The focused beam of light is separated into two beams using a 3-dimensional translation stage and a bifurcated optical fiber. The relative intensity ratio of the two separate beams is adjusted via two additional 3D translation stages before they are detected by the monitor PMT and the test PMT, respectively. Each PMT is located in a dark box. During the measurement, a stable output intensity ratio was adjusted to keep the anode of the monitor PMT unsaturated, which was taken as a reference to examine the dynamic range of the PMT under test. The adopted ratio was 1:10 (monitor PMT vs test PMT) and 1:30 for measuring the anode and the DY8 of the test PMT, respectively. Three signals were read out from the two anodes of both PMTs and the DY8 of the test PMT. The anode signals were decoupled from the positive HV on a spare decoupler board from PandaX-II. Then all three signals were digitized by a waveform digitizer (CAEN V1725B, sampling rate 250 MS/s) when it received a trigger signal which was synchronous with the LED light emission. The recorded data were sent to a server for further analysis.

\begin{figure*}[!htbp]
\centering
\includegraphics[width=0.9\textwidth]{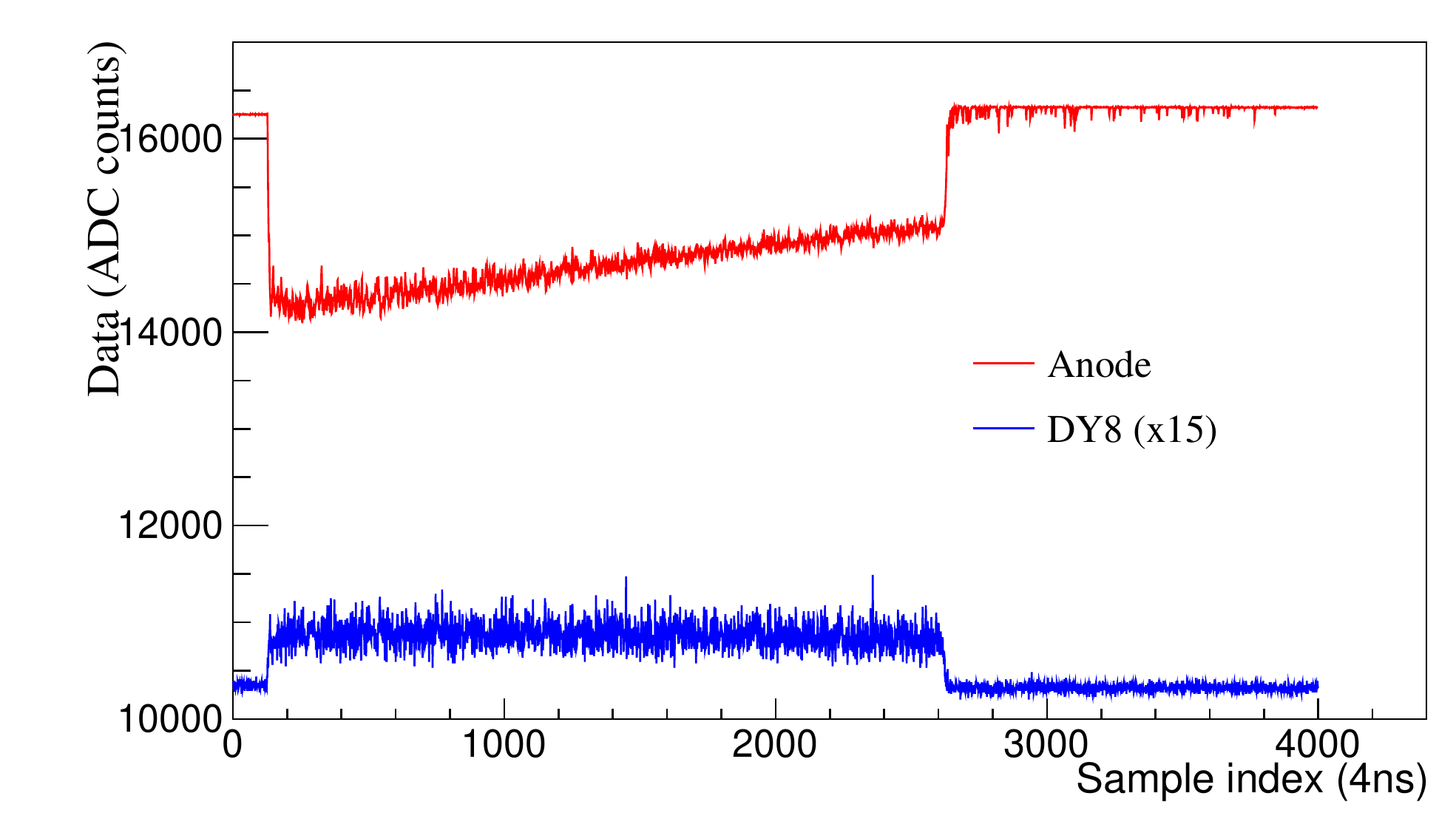}
\caption{An example of two recorded waveforms from the anode and the DY8 of the test PMT, showing that the anode was saturated while the DY8 was unsaturated. In this figure, DY8's signal is scaled by a factor of 15 for better visuality.}
\label{wf}
 \end{figure*}

Figure~\ref{wf} shows an example of two recorded waveforms from the anode and the DY8 of the test PMT. In this example, the amplitude of the signal from the anode was decreasing during the light emission which means the signal was saturated. On the other hand, the signal from DY8 was not saturated. The digitized waveform was then integrated after baseline subtraction. The obtained area should be proportional to the product of the gain and the charge of the PMT if it is not saturated. For the anode, the gain is measured using single photoelectron (SPE) signals in very low-intensity LED calibration runs. When the test PMT is not saturated, the charge ratio between the anode and the DY8 is measured to be 100:1. This ratio is used later to evaluate the charge of the PMT when DY8 is used for the measurement.

\begin{figure*}[htbp]
\centering
\includegraphics[width=0.9\textwidth]{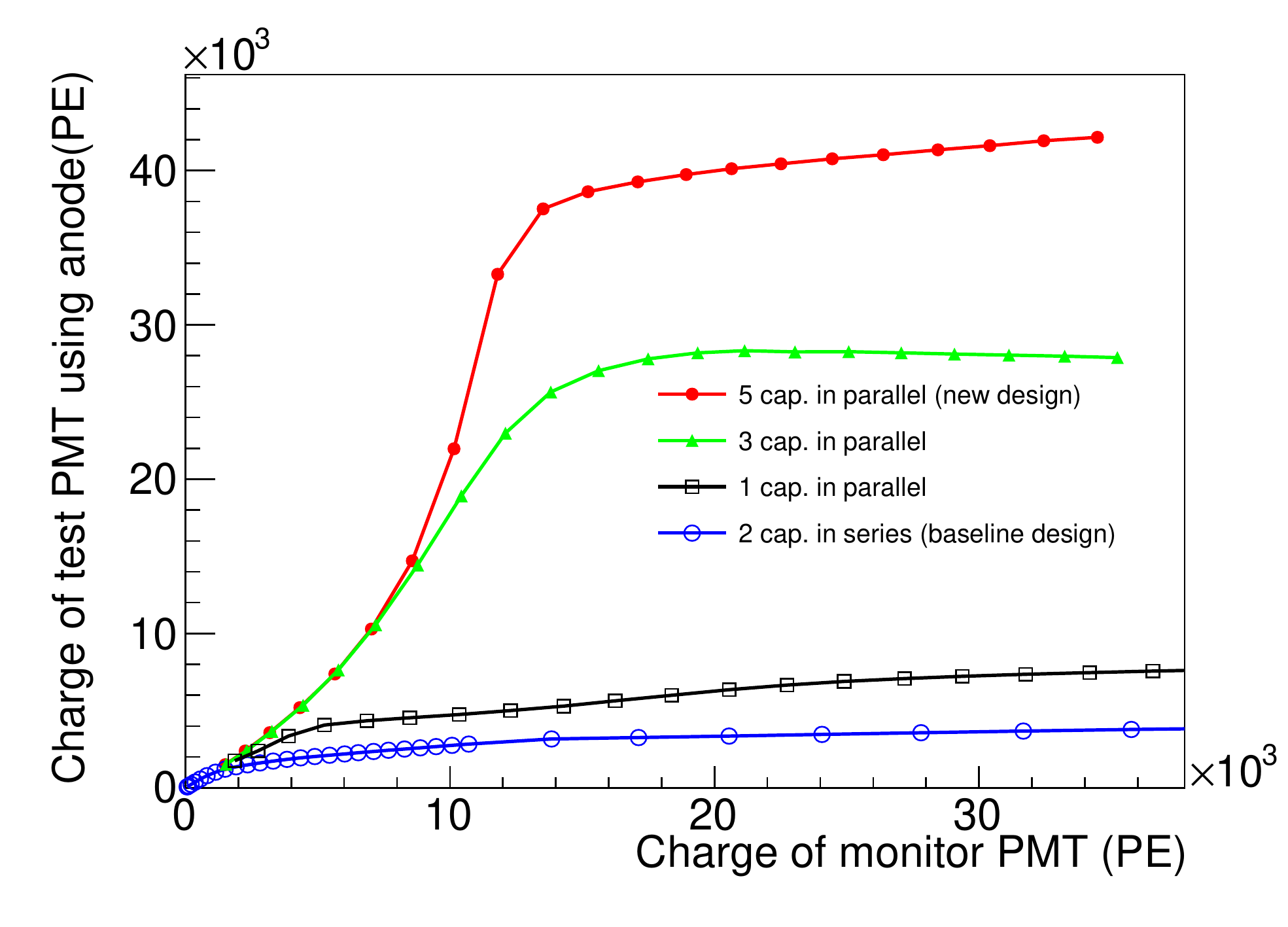}
\caption{Relation between charge of the test PMT using anode and charge of monitor PMT. This includes measurements from the baseline base board and the new design with different capacitor configurations, showing the improvement on the dynamic range if capacitors are connected to ground in parallel rather than in series. Blue curve refers to the baseline design which has two capacitors connected in series to ground. In the new design, we compared the performance of different numbers of capacitors connected in parallel to ground. One capacitor (C$_{2}$ in Figure~\ref{two_design_b}) , three capacitors (C$_{2}$,C$_{1}$,and C$_{4}$ in Figure~\ref{two_design_b}), and five capacitors (C$_{2}$,C$_{1}$,C$_{4}$,C$_{5}$, and C$_{6}$ in Figure~\ref{two_design_b}). The configuration with five capacitors is chosen for the new base board. }
\label{cap_compare}
\end{figure*}

\begin{figure*}[!htbp]
\centering
\includegraphics[width=0.9\textwidth]{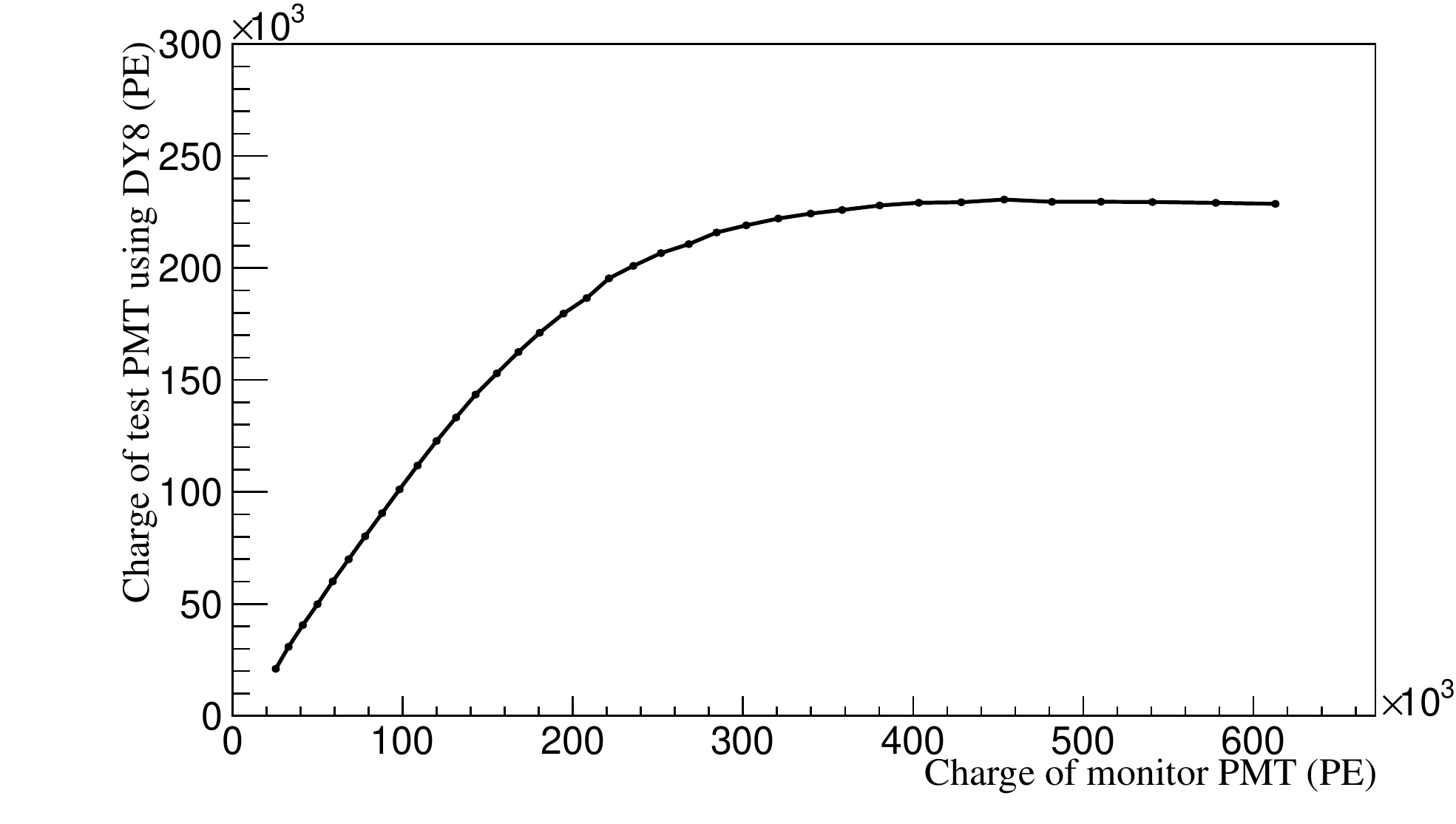}
\caption{Relation between charge of test PMT using DY8 and charge of monitor PMT, showing the dynamic range of the new base board is around 200k PE. Above that the signal from DY8 is also saturated.}
\label{dy8_range}
\end{figure*}

As mentioned above, in the new design of the base board, the number of decoupling capacitors and the way they are connected are different compared to the baseline design. Figure~\ref{cap_compare} shows the relation between charge measured by the test PMT using anode and charge measured by the monitor PMT. The anode of test PMT with the baseline base board start to saturate when the incident light is above 1000 PE and gradually reaches to full saturation. The fully saturated charge is about 4k PE. When the capacitors are connected in parallel, the full saturation happens much later. For the five-capacitor and three-capacitor versions, the anode current starts to deviate upward from the linearity at a certain current level and gradually reaches saturation as the incident light level increases. This somewhat counter-intuitive turning behavior is consistent with that described in Ref.~\cite{ref:hamamatsu}, that the increase of amplification in earlier dynode stages due to voltage redistribution overcomes the decrease of secondary emission ratios in final stages. The fully saturated charge of the five-capacitor new design is about 40k PE, which is 10 times larger than that of the baseline design. Figure~\ref{dy8_range} shows that dynamic range of the test PMT can be further extended if the measurement from DY8 is used. The dynamic range can reach about 200k PE, but with some noticeable level of nonlinearity between 150k PE and 200k PE, which can nevertheless be corrected using real data. This is the equivalent number of PEs expected from the PMT with the maximum charge for a 3.5 MeV energy deposition in the liquid xenon. Thus this range includes the most region of interest for $0\nu\beta\beta$ signals.

\section{Summary}
\label{discussion}
\begin{figure*}[h]
\centering
\includegraphics[width=0.8\textwidth]{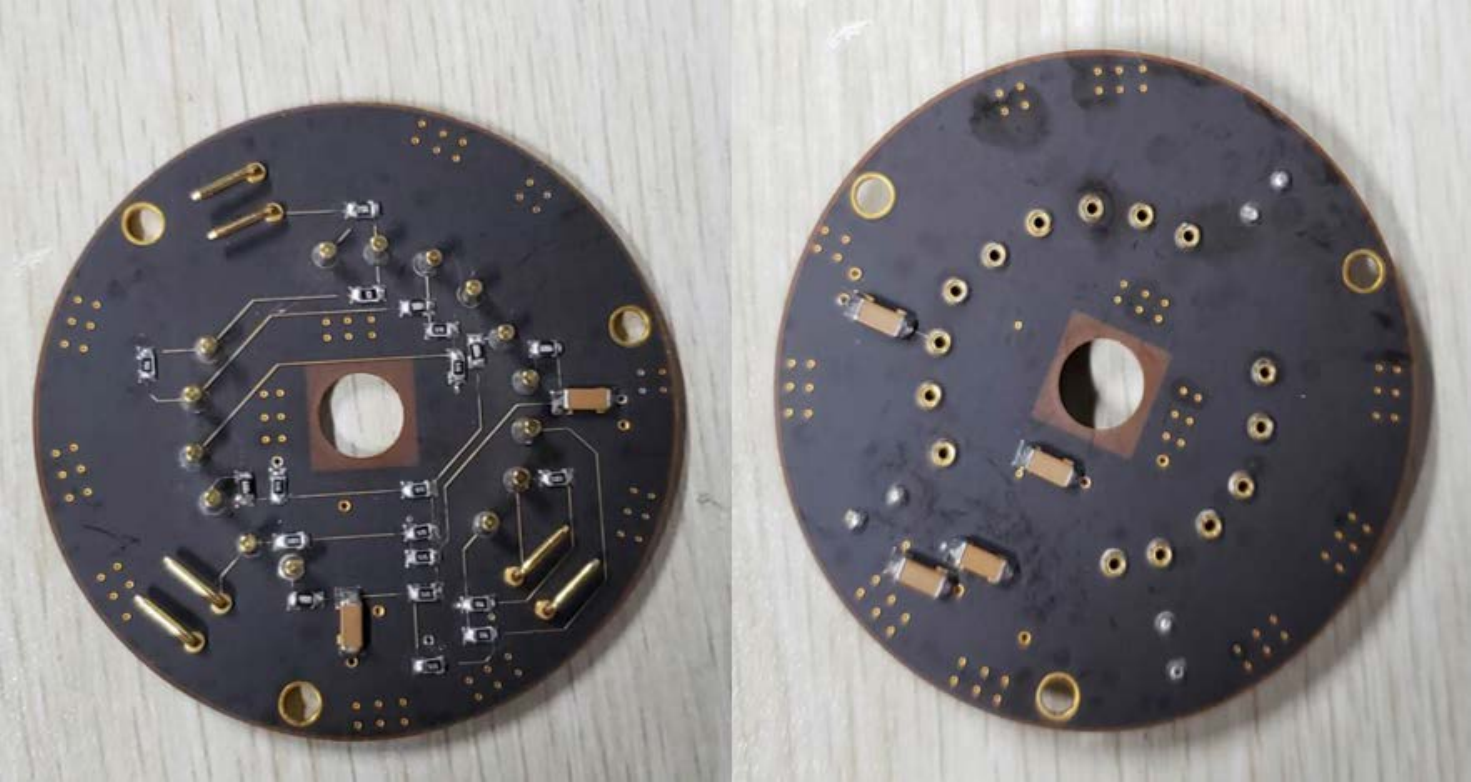}
\caption{Photos of the new base board for 3-inch PMTs Hamamatsu R11410. Left, top view. Right, bottom view.}
\label{structure}
\end{figure*}

Future PandaX DM experiments will be more sensitive in searching for both DM and $0\nu\beta\beta$ signals in the same liquid xenon detector. In PandaX-4T experiment, this requires the dynamic range of the PMT to be at least 150k PE, which is beyond the range provided by the readout base board of the baseline design. In this paper, we have presented an improved design of the readout base board. The new design uses a new configuration of decoupling capacitors. In addition, signals can be read out both from the anode and the DY8. When DY8 is used, the dynamic range of the new base board is measured to be 200k PE which meets the requirement of PandaX-4T. A few new base boards (see photos in Figure~\ref{structure}) have been assembled in the central zone of the top and bottom PMT array of the detector. Their performance will be evaluated using real data from PandaX-4T.
 
\section{Acknowledgement}
This project is supported by grants from the Ministry of Science and Technology of China (No. 2016YFA0400301 and 2016YFA0400302), a Double Top-class grant from Shanghai Jiao Tong University, grants from National Science Foundation of China (Nos. 11875190, 11505112, 11775142 and 11755001), supports from the Office of Science and Technology, Shanghai Municipal Government (18JC1410200), and support also from the Key Laboratory for Particle Physics, Astrophysics and Cosmology, Ministry of Education. This work is supported also by the Chinese Academy of Sciences Center for Excellence in Particle Physics (CCEPP). We thank Changqing Feng at University of Science and Technology of China for useful discussions.


\end{document}